\def\aa{{\,\textrm{\AA}}}

\pdfoutput=1

\documentclass[twocolumn,superscriptaddress,preprintnumbers,amsmath,amssymb,prb]{revtex4}

\usepackage{graphicx}
\usepackage{epstopdf}
\usepackage{setspace}

\usepackage{subfigure}
\usepackage{dcolumn}
\usepackage{bm}
\usepackage{color}

\begin{document}

\title{ \bf{Using Light-Switching Molecules to Modulate Charge Mobility in a Quantum Dot Array}}

\author{Iek-Heng Chu} 
\affiliation{Department of Physics and Quantum Theory Project, University of Florida, Gainesville, Florida 32611, United States}
\author{Jonathan Trinastic}
\affiliation{Department of Physics and Quantum Theory Project, University of Florida, Gainesville, Florida 32611, United States}
\author{Lin-Wang Wang}
\affiliation{Material Science Division, Lawrence Berkeley National Laboratory, One Cyclotron Road, Mail Stop 66, Berkeley, CA 94720}
\author{Hai-Ping Cheng}
\email[Corresponding author: Hai-Ping Cheng,  ]{Email: cheng@qtp.ufl.edu; Tel: 352-392-6256}
\affiliation{Department of Physics and Quantum Theory Project, University of Florida, Gainesville, Florida 32611, United States}

\begin{abstract}
{We have studied the electron hopping in a two-CdSe quantum dot system linked by an azobenzene-derived light-switching molecule. This system can be considered as a prototype of a QD supercrystal. Following the computational strategies given in our recent work [Chu \textit{et al.} J. Phys. Chem. C 115, 21409 (2011)], we have investigated the effects of molecular attachment, molecular isomer (\textit{trans} and \textit{cis}) and QD size on electron hopping rate using Marcus theory. Our results indicate that molecular attachment has a large impact on the system for both isomers.  In the most energetically favorable attachment, the \textit{cis} isomer provides significantly greater coupling between the two QDs and hence the electron hopping rate is greater compared to the \textit{trans} isomer.  As a result, the carrier mobility of the QD array in the low carrier density, weak external electric field regime is several orders of magnitude higher in the $\textit{cis}$ compared to the $\textit{trans}$ configuration.  This is the first demonstration of mobility modulation using QDs and azobenzene that could lead to a new type of switching device.
}

\end{abstract}

\maketitle

\section{Introduction}

Colloidal quantum dots (QDs) have garnered great interest in recent years as potential materials for optoelectronic devices due to the tunability of their electronic, optical, and magnetic properties~\cite{talapin2009prospects}.  Applications for QD nanostructures include solar cells~\cite{Gur21102005}, light-emitting diodes (LED)~\cite{bendall2010,wood2010colloidal}, and thermoelectrics~\cite{wang_thermopower}.  Although early synthesis of these structures yielded disorganized combinations of QDs and organic molecules, more recent advances in ordered self-assembly have allowed for fine control of structural and electronic properties to make functionalized nanostructures~\cite{konstantatos2006ultrasensitive,Yu23052003,Alivisatos16021996}.  In any of the above applications,  efficient charge transport is essential for device performance, which depends on using semiconducting materials with high mobility ($\mu$).  In QD arrays, mobility mainly depends on electronic coupling between QDs through the embedded matrix or interparticle environment~\cite{talapin2009prospects,vanmaekelbergh2005electron}.  Traditional methods of synthesis have used long chains of hydrocarbons as surface ligands that are crucial to control growth and colloidal stability~\cite{talapin2009prospects}.  However, these ligands also act as insulating barriers that limit electron mobility~\cite{Yu23052003,morganPhysRevB.66.075339}.  Recent research has shown that replacing long hydrocarbons with shorter capping molecules such as pyridine, n-buylamine, and metal chalcogenide complexes (e.g., Sn$_2$S$_6$), can improve mobility by several orders of magnitude to 1 - 10 cm$^2$V$^{-1}$s$^{-1}$ due to reduced interdot spacing~\cite{lee2011band,Gur21102005,konstantatos2006ultrasensitive}.  

This recent evidence that mobility is strongly dependent on QD ligands and interdot spacing suggests that conductivity could be modulated by reversible switching of the connectivity between QDs in an array.  One method of achieving this is through the use of photoswitchable ligands that link the QDs.   The photoswitching molecule azobenzene (AB) is an ideal candidate that has been widely researched in applications such as as light-sensitive molecular switches~\cite{Hugel10052002,PhysRevLett.92.158301,del2007tuning,yw-azob} and reversible data storage~\cite{hagen2001photoaddressable}.  Consisting of two benzene rings connected by a double nitrogen linker, the ground state $\textit{trans}$ configuration isomerizes to the $\textit{cis}$ configuration upon exposure to 365~nm wavelength light.  The $\textit{cis}$ configuration can then isomerize back to $\textit{trans}$ due to thermal relaxation or upon 420~nm wavelength light irradiation.  Previous computational studies using azobenzene in metal-AB-metal~\cite{PhysRevLett.92.158301,yw-azob} and CNT-AB-CNT~\cite{del2007tuning} junctions have demonstrated distinct conductance patterns when switching between the $\textit{trans}$ and $\textit{cis}$ configurations~\cite{mativetsky_azo,PhysRevLett.92.158301}.

Since mobility in QDs is dependent upon tunneling barriers and interdot spacing, a similar mechanism as described in metal-molecule junctions could be applied to QD arrays to reversibly switch tunneling barrier lengths.  Azobenzene has already been synthesized with QDs, either as capping ligands on CdS QDs~\cite{fang2010temperature} or as a way to reversibly assemble and disassemble a suprastructure of gold nanoparticles~\cite{Klajn19062007}.  However, no experimental or theoretical research has investigated the questions of mobility and conductance in QDs linked by reversible molecules such as azobenzene.

To investigate charge transport in such a system, the charge hopping process between AB-linked QDs is of fundamental importance. In our recent work, we have studied different possible mechanisms for the charge hopping in a two-CdSe QD system\cite{doi:10.1021/jp206526s}. We have found that the multi-phonon mechanism, which can be described by Marcus theory~\cite{marcus}, dominates the hopping process. In this case, excited electrons carrying higher kinetic energy first rapidly lose energy to phonons and relax to the conduction band edge state~\cite{relax1,relax2}. This relaxation process completes prior to hopping between QDs. Thus, by studying hopping between these band edge states, one can gain physical insight into the relevant transport processes.

In this work, we have studied the electron hopping rate in a simplified system that contains two CdSe QDs linked by an AB derivative.  This process serves as the primary step for electron transport in an actual QD array and is therefore of fundamental significance. We have used CdSe QDs in the present study because they are widely used in experiment and demonstrate relatively high conductivities~\cite{talapin2009prospects,Kovalenko12062009}. From the electron hopping rate, we have estimated the carrier mobility in the array that can be compared to future experimental results. The article is organized as follows: we give a brief description of our model system and the computational methods in Section II, present results in Section III, and end with brief conclusions  in Section IV.

\begin{figure}
\includegraphics[width=7.5cm]{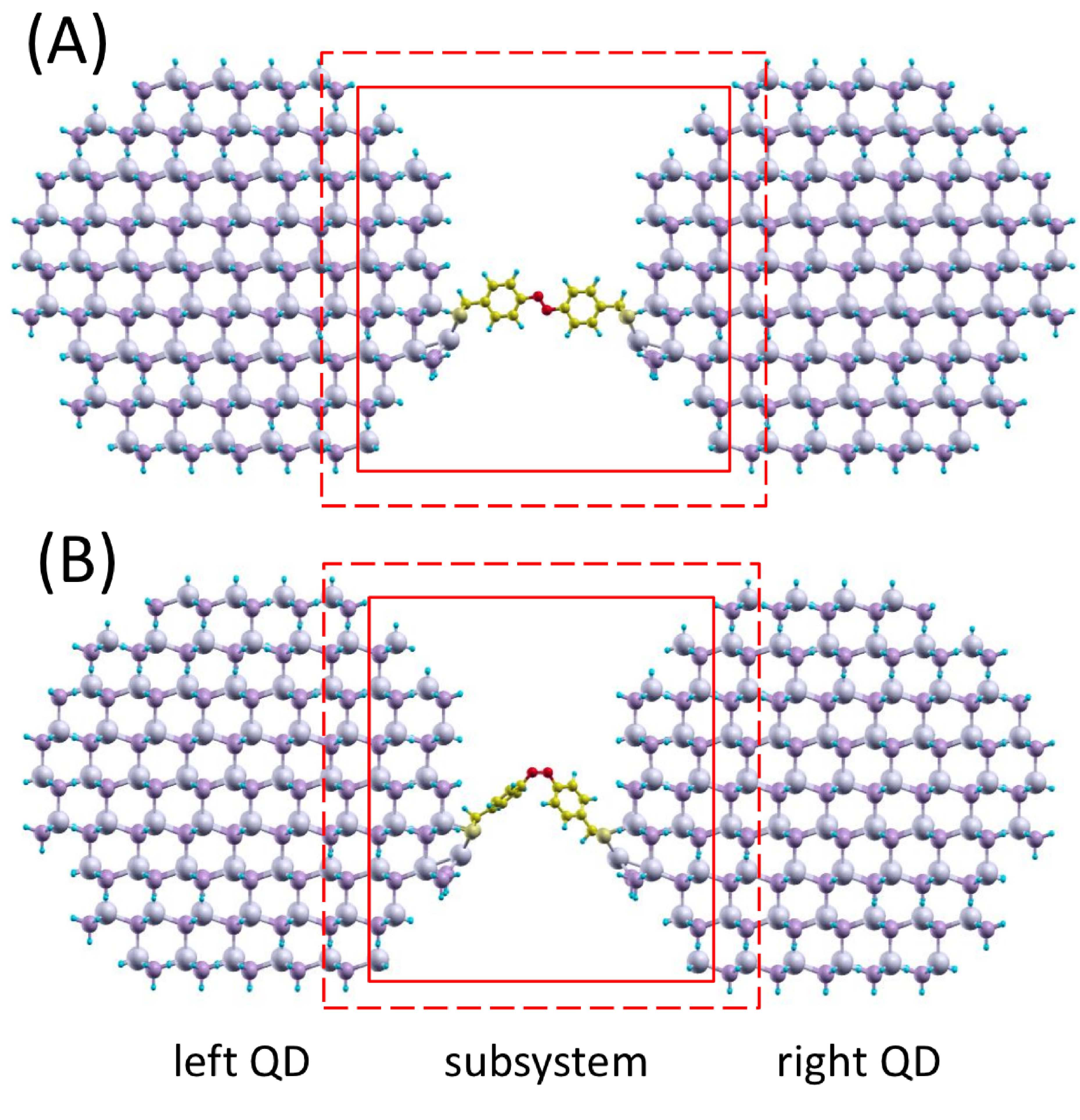}
\caption{(color online) Our model system contains two CdSe QDs convalently linked by an (A) $\textit{trans}$ and (B) $\textit{cis}$ azobenzene derivatives. The subsystem is cast out from the two QDs, as indicated by the red dashed line.   The space between the dashed and solid red lines indicates the region in which the mask function \textit{M(\textbf{r})} is applied for the charge patching method (see text for details).  This figure is created using XCrysden~\cite{xcrysden}. }\label{total_sys}
\end{figure}

\section{Model and Computational Methods}
Our model system contains two CdSe QDs covalently linked by the AB derivative with either the $\textit{trans}$ or $\textit{cis}$ isomer, as shown in Figure~\ref{total_sys}. This molecule contains an AB fragment in the center, with a CS$_{2}$ unit linked to each side.  This linker has been chosen because previous experimental studies have already used sulfur-containing molecules such as Sn$_2$S$_6$ to bind to CdSe QDs~\cite{Kovalenko12062009}. The surface of the QDs is passivated by artificial hydrogen atoms to remove gap states due to dangling bonds. Within density functional theory (DFT), the electronic structure can be studied provided that the total charge density is computed. However, such a system usually contains thousands of atoms, which makes a conventional DFT calculation impractical. Here, we have calculated the total charge density of the systems by applying computational methods used in our previous work~\cite{doi:10.1021/jp206526s} and briefly described below.

\subsection{Molecular attachment calculations}
To determine the most energetically favorable molecular attachments of the AB derviative connecting the two CdSe QDs, we have investigated the molecular attachment between two flat CdSe (10-10) surfaces, which are approximated as the QD surfaces. Such an approximation is justified by the fact that the actual molecular attachment involves only a very small region and that the (10-10) surface is the most stable surface in a CdSe crystal~\cite{doi:10.1021/jp0445573}. Each QD surface is then modeled by a three-layer slab covalently linked to the AB derivative. On each CdSe surface, a standout Cd atom is added near the AB derivative to satisfy the local electron counting rule. The surfaces are periodic in the $x$ and $y$ directions  and we have set the intermolecular distance to be about 10~$\aa$ to minimize the interactions between periodic images of the molecules. A 10 $\aa$ vacuum layer is added above each slab along the $z$-direction to avoid nonphysical interactions between the slabs and their images. 

 The structural optimizations have been performed using the plane-wave DFT code VASP~\cite{PhysRevB.47.558,PhysRevB.54.11169}. The LDA exchange-correlation functionals~\cite{PhysRevLett.45.566}, together with a kinetic energy cutoff of 440 eV have been used within the projected augmented wave method~\cite{PhysRevB.50.17953}. Several initial attachments have been used and then the corresponding atomic structures have been optimized. Since the total number of atoms of the system may differ among the initial attachments and the two isomers, the bonding energy calculations have been carried out to determine the most stable molecular attachments. 

\subsection{Construction of the total charge density with \textit{ab initio} accuracy}
To construct the total charge density of the model system, we have adopted the divide-and-conquer method.  This procedure allows for the electronic structure calculation of our large model system (two QDs plus AB derivative) by combining separate calculations of the charge density in the molecule-QD attachment region and the single-QD regions.  We have first constructed a subsystem that includes the AB derivative and nearby QD regions bounded by the dashed red lines in Figure~\ref{total_sys}. We have again used artificial hydrogen atoms to passivate the QD surfaces.  Across all sizes of QD used in this study, this subsystem contains only a few hundred atoms and thus its charge density can be obtained by a standard DFT calculation. In the present work, such calculations have been performed using the plane-wave DFT code PEtot~\cite{petot} along with LDA exchange correlation functionals and norm-conserving pseudopotentials~\cite{NCPP}.The kinetic energy cutoff has been chosen to be 62 Ry (about 840 eV) to ensure total energy convergence.

Next, the charge densities in each QD region have been calculated separately using the well-tested charge patching method (CPM)~\cite{PhysRevLett.88.256402}, in which the total charge density of a large system can be determined by assuming that the charge density at a given point in real space only depends on the neighboring atoms.  This allows for the calculation of charge density in a small region that can be extended to describe large systems such as the QDs used here.  This method has been shown to give a charge density differing from a self-consistent DFT counterpart by only 0.1 $\%$.   Finally, to connect the charge densities calculated for the subsystem ($\rho_{sub}(\bf{r})$) and the two QDs ($\rho_{QD}(\bf{r})$), we have introduced a mask function $M({\bf r})$ that smoothly varies from 0 to 1 when ${\bf r}$ crosses from the QD side (red dashed lines) to the central molecular side (red solid lines), as illustrated in Figure \ref{total_sys}. The total charge density ($\rho_{tot}(\bf{r})$) can then be expressed as

\begin{equation}
\rho_{tot}(\bf{r})=\it{M}(\bf{r})\rho_{sub}(\bf{r})+[1-\it{M}(\bf{r})]\rho_{QD}({\bf r})\label{tot_den}
\end{equation}

 It is important to note that to obtain the correct total charge, one might need to rescale $\rho_{sub}({\bf r})$ in the subsystem slightly.

Once the total charge density has been calculated, the total DFT potential $V_{tot}({\bf r})$ can be solved directly from the Poisson equation. Then one arrives at the single-particle Kohn-Sham (KS) equation~\cite{KS} given as:

\begin{eqnarray}
&& H({\bf r})\psi_i({\bf r})=-\frac{\hbar^2}{2m}\nabla^{2}+V_{eff}({\bf r})=\epsilon_{i}\psi_i({\bf r}),\\
&& V_{eff}({\bf r})=V_{tot}({\bf r})+\widehat{V}_{NL}
\end{eqnarray}

where $\widehat{V}_{NL}$ is the nonlocal part of the atomic pseudopotentials; $\psi_i({\bf r})$ is the \textit{i}$^{th}$ eigen-state with the state energy $\epsilon_i$. Instead of directly solving the above equation, we have applied the folded spectrum method (FSM)~\cite{FSM} to obtain the conduction band edge states. These are the primary states of interest when calculating electron mobility because it is assumed that excited electrons lose energy to phonons and relax to the band edge on a timescale much smaller compared to the hopping rate.   In particular, we have solved the equation $[H({\bf r})-\epsilon_{ref}]^2\psi_i({\bf r})=(\epsilon_{i}-\epsilon_{ref})^2\psi_i({\bf r})$ where $\epsilon_{ref}$ is some energy reference inside the band gap and is close to the conduction band minimum. Consequently, the first few states solved from this equation correspond to those closest to the conduction band edge.   Here, the energy reference is estimated using the generalized moments method~\cite{GMM}. In this work, since we have only studied the electron hopping between conduction band states, we do not expect the well-known DFT band gap problem to significantly affect our results. 

\subsection{Electron hopping theory}
The electron hopping from QD a to QD b is dominated by the multi-phonon process, which can be described by the Marcus hopping rate ($k_{ab}$) equation~\cite{marcus}:

\begin{equation}
k_{ab}=|V_{ab}|^2\sqrt{\frac{\pi}{\lambda k_BT\hbar^2}}\exp[-(\lambda+\epsilon_a-\epsilon_b)^2/4\lambda k_BT]\label{MT}.
\end{equation}
,
Here, state $\psi_{a(b)}({\bf r})$ is localized on the QD a(b). $V_{ab}$ is the electronic coupling between states $\psi_{a}({\bf r})$ and $\psi_{b}({\bf r})$ which can be calculated by their energy anticrossing (AC)~\cite{AC} when one state is under perturbation by an external potential. $\lambda$ is the reorganization energy, which corresponds to the atomic relaxation energy after the electron hops from QD a to QD b. $\epsilon_{a(b)}$ is the state energy when an electron occupies QD a(b). When the two QDs are the same size, $\epsilon_{a}=\epsilon_{b}$. $k_B$ and $T$ are Boltzmann's constant and temperature, respectively.  This theory has been proven to be in very good agreement with a more advanced treatment in which the atomic motions are considered quantum mechanically within the harmonic approximation~\cite{Huang22121950,PhysRevB.79.115203}.  Both $V_{ab}$ and $\lambda$ are the central quantities that must be calculated to determine the hopping rate using Marcus theory.

We have first calculated the electronic coupling $V_{ab}$ between the two lowest conduction band states that are involved in electron hopping between the QDs.   Without the presence of the AB linker molecule, the conduction band edge states from each QD would be nearly degenerate.  The inclusion of the linker molecule splits the energy of these two states. The electronic coupling $V_{ab}$ between two QD states can then be computed by adding an artifically small, Gaussian-like potential to one QD to cause an AC curve between the energies of these two QD states. The minimal energy splitting between these two states is the AC energy, equal to $2V_{ab}$~\cite{AC}.

When an electron at one QD hops to another QD, there is an atomic relaxation due to the induced forces. The resulting relaxation energy is known as the reorganization energy ($\lambda$).  To calculate this quantity, we have computed the electron-phonon coupling matrix, defined as $C_{ij}({\bf R})=\langle\psi_i|\partial H/\partial{\bf R}|\psi_{j}\rangle$. The diagonal elements $C_{ii}({\bf R})=\langle\psi_{i}|\partial H/\partial{\bf R}|\psi_{i}\rangle$ correspond to additional forces acting on the atoms at coordinates $\{{\bf R}\}$ due to the occupation change (addition/removal of an electron) at state $\psi_{i}$. Since the QD states $\psi_{a}$ and $\psi_{b}$ are mostly localized on the two QDs instead of the AB molecule, we have approximated this energy as the sum of the relaxation energies in each single QD. The calculation of the coupling matrix elements $C_{aa(bb)}$ is non-trivial as it usually involves thousands of atoms.  To make such a calculation feasible, we have used CPM to calculate the coupling matrix elements. Then the valence force field (VFF)~\cite{VFF} has been applied to the single QD and the relaxation energy due to the induced forces acting on each atom has been obtained. 

\subsection{Calculating the carrier mobility}
After the electron hopping rate is obtained for the QD-AB-QD system, one can calculate the electron mobility in the low carrier density regime with a weak external electric field~\cite{PhysRevB.4.2612, doi:10.1021/nl9021539}. To do this, we have constructed a cubic QD network containing $N_{x}\times N_{y}\times N_{z}$  QDs with $N_{x}=N_{y}=N_{z}=N$  (e.g. $N=50$). In this network, all the QDs are the same size (characterized by the QD diameter D), and any two adjacent QDs (labeled as $a$ and $b$) are connected by the AB derivative, with the hopping rate from $a$ ($b$) to $b$ ($a$) being $k_{ab (ba)}$. Then the conductance between them is given by~\cite{doi:10.1021/nl9021539} $G_{ab}=G_{ba}=e^2n_{a}k_{ab}/k_{B}T=e^2n_{b}k_{ba}/k_{B}T$. Here, $n_{a(b)}$ is the occupation number of state $\psi_{a (b)}$ in the QD $a(b)$; $e$ and $k_{B}$ are the elementary charge and Boltzmann constant, respectively. By applying an external electric field along certain direction, e.g. the $z$-direction,  we have then solved a set of linear equations $\sum_{i}(V_{i}-V_{j})G_{ij}=0$ to obtain the potential at each QD site in the network. Then we have calculated the total current $I_{z}$ and hence the total conductance $G_{z}$. In the low carrier density regime, the electron mobility can be written as  

\begin{equation}
\mu_{z}=G_{z}/neL_{z}
\end{equation}

with $L_{z}=N_zD$ being the network size along the $z$-direction, and $n$ the average concentration of electron carriers. When there is no fluctuations (size or attachment) in the QD array, we have $n=n_a/D^3$. Now the electron mobility reads:

\begin{equation}
\mu_{z}=\frac{ek_{ab}}{k_BT}D^2
\end{equation}

\section{Results}
\begin{figure}
\includegraphics[width=7.5cm]{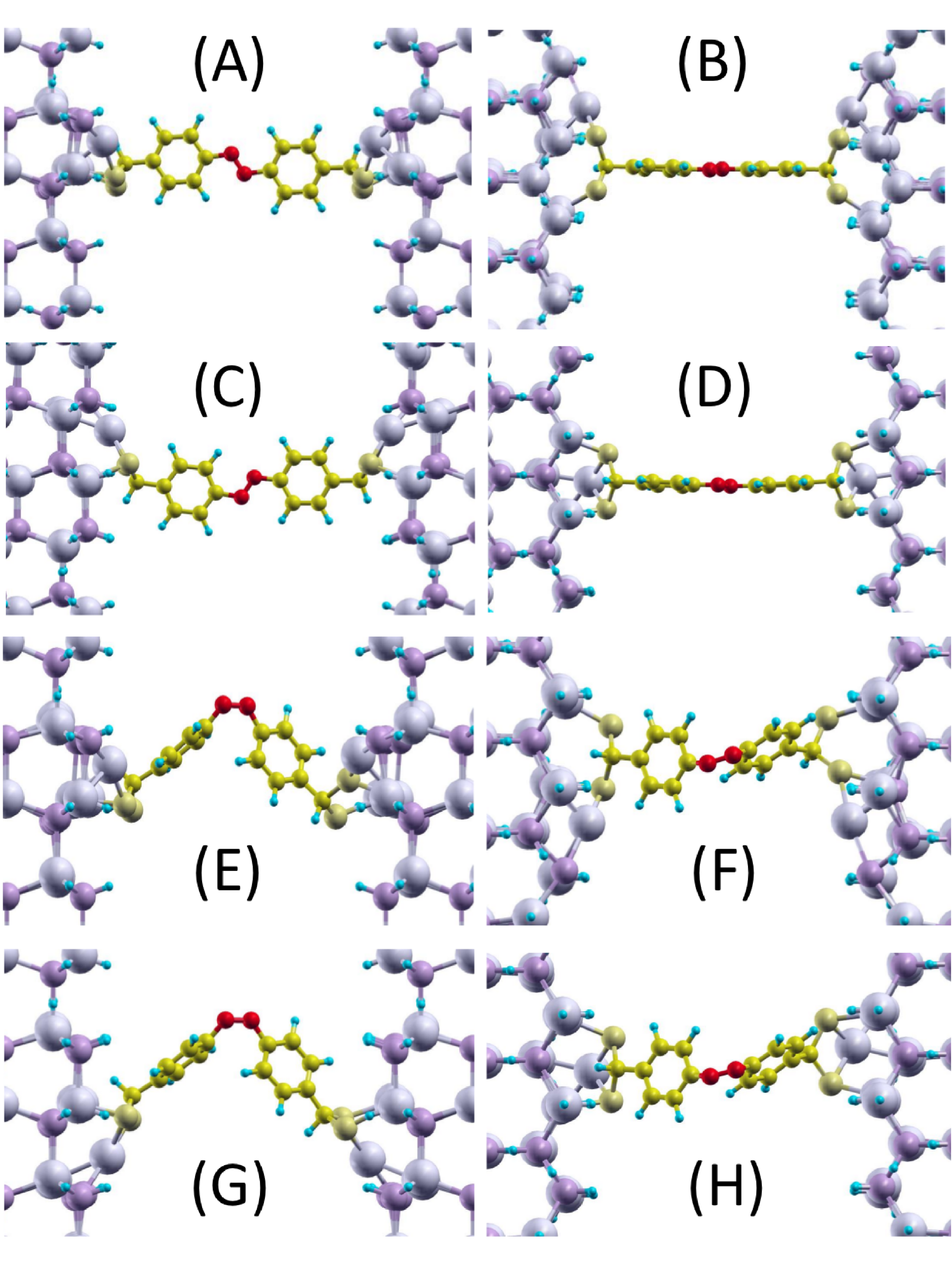}
\caption{(color online) Side views of type I trans- molecular attachment (A,B); type II trans- molecular attachment (C,D); type I cis- molecular attachment (E,F) and type II cis- molecular attachment (G,H).}\label{attachs}
\end{figure}

We have first investigated the optimal attachment of the AB derivative connecting two flat CdSe (10-10) surfaces, in which we have tested different initial attachments (see Sec. II.A). The two most stable attachments for both the \textit{trans} and \textit{cis} isomers are shown in Figure~\ref{attachs}. For the type I attachment (Figure~\ref{attachs}(A,B) for \textit{trans} and (E,F) for \textit{cis}), one of the S atoms is covalently bonded to both the standout Cd atom and the other Cd atom on the surface, whereas the other S atom is only bonded to one Cd atom on the surface. In type II attachment, both S atoms are bonded to both the standout Cd atom and one Cd atom on the surface (Figure~\ref{attachs}(C,D) for \textit{trans} and (G,H) for \textit{cis}). Furthermore, bonding energy calculations show that the type II attachment is more stable as its bonding energy is 0.8 eV higher than the type I attachment. This is due to more S-Cd chemical bonds between the linker molecule and the CdSe surface.  For each AB isomer, these two types of attachment give very similar surface-surface distances, and hence similar QD-QD distance. In particular, the surface-surface distance is around 17$\aa$ for the \textit{trans} isomer while it is about 14$\aa$ for the \textit{cis} isomer. The following results discuss electron hopping rates and carrier mobility for both attachment types to investigate the effect of bonding type on transport. 

We have then constructed the QD-AB-QD systems for both molecular isomers using four different sizes of the CdSe QD. The diameter ($D$) and number of atoms ($N_a$) in each single QD are given in Table I. Such large systems are usually formidable for a direct DFT calculation. Therefore, we have followed the strategy given in Sec. II.B to construct the total charge density of each system.

\begin{table*}[t]
\caption{The electronic coupling $V_{ab}$ for molecular attachment types I and II in the \textit{trans} and \textit{cis} isomers. $N_a$, $D$ and $\lambda$ are the number of atoms in a single QD, the QD diameter and the reorganization energy, respectively.}
\begin{tabular}{ c c c c c c c}
\hline \hline
\noalign{\medskip}
$\quad$ $N_a$ \quad & $\quad$ $D$~(nm) \quad & $\quad$ $\lambda$~(meV) \quad & $\quad$ $V^{cis,I}_{ab}$(meV) & $\quad$ $V^{cis,II}_{ab}$(meV) & $\quad$ $V^{trans,I}_{ab}$(meV) & $\quad$ $V^{trans,II}_{ab}$(meV)\\
\noalign{\medskip}
\hline   
\noalign{\medskip}
 \quad 272 \quad & \quad 1.9 \quad  & \quad 215 \quad & \quad 0.31 & \quad 1.95 & \quad 0.12 & \quad 0.09 \\
 \quad 460 \quad & \quad 2.4 \quad & \quad 136 \quad & \quad 0.02 & \quad 1.68 & \quad 0.05 & \quad 0.08 \\ 
 \quad 1051 \quad & \quad 3.4 \quad & \quad 62 \quad & \quad 0.03 & \quad 0.31 & \quad $\le$~0.01 & \quad $\le$~0.01  \\
 \quad 1916 \quad & \quad 4.3 \quad & \quad 32 \quad & \quad 0.02 & \quad 0.04 & \quad $\le$~0.01 & \quad $\le$~0.01 \\
 \noalign{\medskip}
 \hline \hline
\end{tabular}
\label{tab1}
\end{table*}

Once the total charge densities for the subsystems using both $\textit{trans}$ and $\textit{cis}$ molecular isomers have been found, we have obtained the total potentials $V({\bf r})$ by solving the Poisson equation. We have then computed the conduction band (CB) edge states utilizing the FSM. Based on our results, there are always two states at the CB edge, each of which are localized on one of the two QDs (We call them CBM and CBM+1). For all systems studied, the overlap between these two states is small but depends on the interdot distance and plays a crucial role in determining electron hopping rates.  In QD systems, when an excited electron initially occupies one of these CB states (e.g., CBM+1) localized on one QD and hops to the other state (CBM), an electron transfer between the QDs can occur. This is the first step for electron transport through a QD array in the experiment. As described in Sec. II.C, we determine the hopping rate using Marcus theory, which requires the calculation of two system quantities: (i) electronic coupling $V_{ab}$ between CBM and CBM+1 within the band AC picture and (ii) the reorganization energy $\lambda$ due to the electron-phonon interaction.

First, we have computed the coupling constant $V_{ab}$ for $\textit{trans}$ and $\textit{cis}$ isomers comparing attachment types I and II. The coupling constants across all QD sizes and for each attachment type are given in Table I.  All the calculated coupling constants are less than 2 meV, which is consistent with our findings that the overlap between the two QD states is very small.  In the cases of the 1051- and 1916-atom QDs with the \textit{trans} isomer, we have used 0.01 meV as the upper bound for the coupling due to extremely small calculated values.  The subsequent hopping rates and carrier mobility calculated using this coupling value can be considered as upper bounds (see Table II and III).  The molecular attachment type has a large effect on the couplings in both isomers. In type II attachment, which is more energetically stable, the coupling values are much higher in the \textit{cis} isomer than in the \textit{trans} isomer. This large coupling difference is likely due to the fact that the QD-QD distance when using the $\textit{trans}$ linker molecule is about 3 $\AA$ longer than when using the $\textit{cis}$ linker molecule.  The electronic coupling typically decays exponentially as a function of interdot distance~\cite{transport_review}, which relates to the overlap of the two QD states. Therefore, the $\textit{trans}$ electronic couplings are very small compared to those in the \textit{cis} isomer. In the case of the $\textit{cis}$ isomer, the electronic coupling is always larger for the type II compared to the type I attachment. This is because the type II attachment is more stable, since more bonds exist between the QD and the linker molecule that enhance the coupling.

The coupling constant also depends on the size of the QD for both $\textit{trans}$ and $\textit{cis}$ isomers. In general, the electronic coupling decreases as the QD diameter increases. This relates to the fact that smaller QDs have stronger quantum confinement effects. This results in the QD conduction band energies moving upward, which reduces the potential barrier between the QDs~\cite{doi:10.1021/jp206526s}. Another reason is that the state has higher amplitude at the QD surface for smaller QDs, which increases the coupling. This size dependence is strongest when using the $\textit{cis}$ linker molecule with the type II attachment, in which the coupling is increased by a factor of 7 when the diameter is reduced from 4.3 nm to 3.4 nm; further reductions in the QD size have a subsequent increase in the electronic coupling. It is important to note that this difference in coupling due to QD size diminishes for the smallest QDs.  Specifically, when the QD diameter reduces from 2.4 nm to 1.9 nm, the resulting electronic coupling change is less than 20$\%$. 

To calculate the reorganization energy of the system, as discussed before, we have computed the induced forces acting on the atoms at coordinates $\{{\bf R}\}$, as defined in Sec.~II.C. We have calculated this quantity for the 272-atom and 460-atom QDs, whereas we have adopted the values for the 1051-atom and 1916-atom QDs from our previous work~\cite{doi:10.1021/jp206526s}, given in Table I. The reorganization energy is much larger than the electronic coupling, and it linearly scales with the inverse of QD size, which has been proven in our previous work~\cite{doi:10.1021/jp206526s}. 

\begin{figure}
\includegraphics[width=7.5cm]{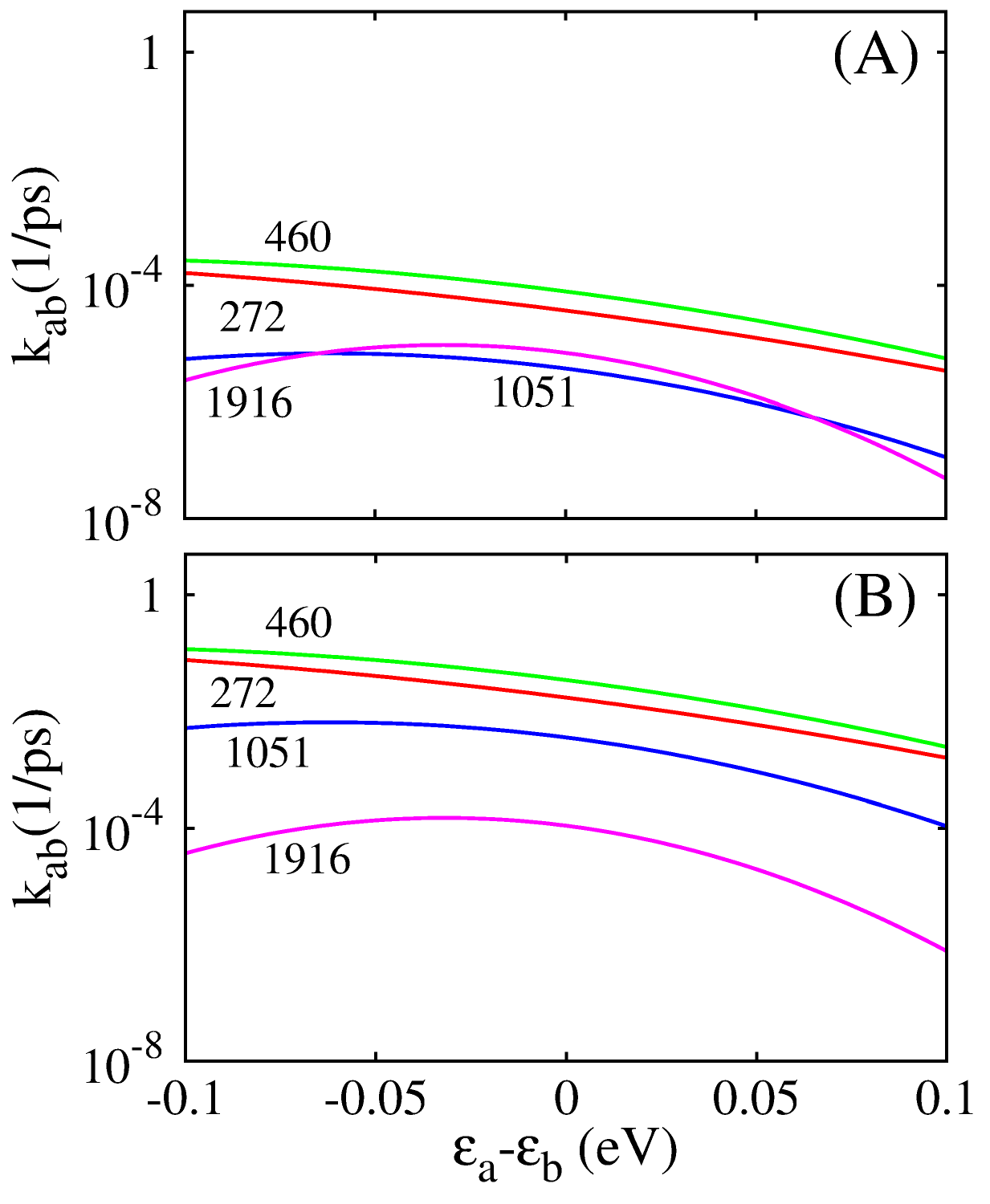}

\caption{(color online) Marcus electron hopping rate $k_{ab}$ vs. state energy difference $\epsilon_a-\epsilon_b$ for the two-QD system with type II attachment using (A) \textit{trans} and (B) \textit{cis} isomer. In the 1051- and 1916-atom QD cases with the \textit{trans} isomer, the upper bound of the electronic coupling (0.01 meV) is used.}\label{rate_eab}
\end{figure}

\begin{table*}[t]
\caption{ The Marcus electron hopping rate $k$ at $\epsilon_a-\epsilon_b=0$ and the carrier mobility $\mu$ in the linear regime using different sizes of the QD and molecular \textit{trans} and \textit{cis} isomers, in type I molecular attachment.  In the cases of $D=$3.4 nm and 4.3 nm with the \textit{trans} isomer, the upper bound of the electronic coupling, 0.01 meV, is used. The corresponding $k$ and $\mu$ can be considered as their upper bounds. }
\begin{tabular}{ c c c c c}
\hline \hline
\noalign{\medskip}
  \quad $D$ (nm) \quad & \quad $k_{cis}$ (1/ps) \quad & \quad $k_{trans}$ (1/ps) \quad & \quad $\mu_{cis}$ (cm$^2$/V/s) \quad & \quad $\mu_{trans}$ (cm$^2$/V/s) \quad \\
\noalign{\medskip}
\hline   
\noalign{\medskip}
 \quad 1.9 \quad & \quad $4.3\times$10$^{-4}$ & \quad $6.5\times$10$^{-5}$ & \quad $6.1\times$10$^{-4}$ & \quad $9.1\times$10$^{-5}$ \quad \\
 \quad 2.4 \quad & \quad $4.9\times$10$^{-6}$ & \quad $3.0\times$10$^{-5}$ & \quad $1.1\times$10$^{-5}$ & \quad $6.8\times$10$^{-5}$ \quad \\ 
 \quad 3.4 \quad & \quad $3.3\times$10$^{-5}$ & \quad $\le 3.7\times$10$^{-6}$ & \quad $1.5\times$10$^{-4}$ & \quad $\le 1.7\times$10$^{-5}$ \quad \\
 \quad 4.3 \quad & \quad $2.7\times$10$^{-5}$ & \quad $\le 6.9\times$10$^{-6}$ & \quad $2.0\times$10$^{-4}$ & \quad $\le 4.9\times$10$^{-5}$ \quad \\
 \noalign{\medskip}
 \hline \hline
\end{tabular}
\label{tab2}
\end{table*}

\begin{table*}[t]
\caption{ The Marcus electron hopping rate $k$ at $\epsilon_a-\epsilon_b=0$  and the carrier mobility $\mu$ in the linear regime using different sizes of the QD and molecular \textit{trans} and \textit{cis} isomers, in type II molecular attachment. In the cases of $D=$3.4 nm and 4.3 nm with the \textit{trans} isomer, the upper bound of the electronic coupling, 0.01 meV, is used. The corresponding $k$ and $\mu$ can be considered as their upper bounds. }
\begin{tabular}{ c c c c c}
\hline \hline
\noalign{\medskip}
  \quad $D$ (nm) \quad & \quad $k_{cis}$ (1/ps) \quad & \quad $k_{trans}$ (1/ps) \quad & \quad $\mu_{cis}$ (cm$^2$/V/s) \quad & \quad $\mu_{trans}$ (cm$^2$/V/s) \quad \\
\noalign{\medskip}
\hline   
\noalign{\medskip}
 \quad 1.9 \quad & \quad $1.7\times$10$^{-2}$ & \quad $3.7\times$10$^{-5}$ & \quad $2.4\times$10$^{-2}$ & \quad $5.1\times$10$^{-5}$ \quad \\
 \quad 2.4 \quad & \quad $3.4\times$10$^{-2}$ & \quad $7.8\times$10$^{-5}$ & \quad $7.7\times$10$^{-2}$ & \quad $1.7\times$10$^{-4}$ \quad \\ 
 \quad 3.4 \quad & \quad $3.5\times$10$^{-3}$ & \quad $\le 3.7\times$10$^{-6}$ & \quad $1.6\times$10$^{-2}$ & \quad $\le 1.7\times$10$^{-5}$ \quad \\
 \quad 4.3 \quad & \quad $1.1\times$10$^{-4}$ & \quad $\le 6.9\times$10$^{-6}$ & \quad $7.9\times$10$^{-4}$ & \quad $\le 4.9\times$10$^{-5}$ \quad \\
 \noalign{\medskip}
 \hline \hline
\end{tabular}
\label{tab3}
\end{table*}

Using $V_{ab}$ and $\lambda$, we have then calculated the electron hopping rate $k_{ab}$ between the two QDs using Marcus theory, as given in Eq.(\ref{MT}).  When the two QDs in the model system are of exactly the same, their QD state energies are the same, i.e. $\epsilon_a-\epsilon_b=0$. In reality, neighobring QDs in a supercrystal may vary in size.  This size fluctuation leads to small differences in their state energies that influence $k_{ab}$. The hopping rate as a function of the state energy difference is plotted in Figure~\ref{rate_eab} for the type II attachment. The corresponding curves for the type I attachment only differ by an overall scaling factor $|V_{ab}|^2$ from the type II attachment results. Except for the 1916-atom QD case, the hopping rates at $\epsilon_a-\epsilon_b=0$ for the $\textit{cis}$ isomer using the type II attachment, as shown in Table III, are consistently at least an order of magnitude larger than those using the type I attachment, given in Table II. This is due to the much stronger coupling between the two QD states when using the type II attachment. On the other hand, the \textit{trans} isomer leads to similar hopping rates using both attachement types. Across all QD sizes, $k_{ab}$ in the type II attachment is at least an order of magnitude larger for the \textit{cis} compared to \textit{trans} isomer. It is worth pointing out that in the type II attachment, the computed hopping rate in the 272-atom QD case is smaller than that in the 460-atom QD case, regardless of isomer type. This is likely due to the stronger electron-phonon interaction, and hence larger reorganization energy for the 272-atom QD case than for the 460-atom QD case, while the electronic couplings are similar in these two cases for both isomers. Overall, such a significant difference in the hopping rate between the AB $\textit{trans}$ and $\textit{cis}$ isomers indicates that the AB molecule can serve as a molecular switch in an actual QD array. 

Using the calculated electron hopping rates, we have then constructed a cubic QD array and calculate its carrier mobility in the low carrier density regime under a weak electric field as discussed in Sec. II.D. In this work, we assume that the QD arrays are simple cubic and we do not consider the QD size fluctuation, i.e. the QD used in each array is the same size. The mobility values are given in Tables II and III. We have found that, when using the type II attachment, the mobility for each QD size in the $\textit{cis}$ isomer are at least one order of magnitude larger than its counterpart in the $\textit{trans}$ isomer. Except for the 4.3 nm QD case, the $\textit{cis}$ isomer mobility is always at least 500 times larger than the $\textit{trans}$ isomer mobility. Since the size of these QDs are comparable to those synthesized experimentally, our results can be directly compared to future experiment. For type I attachment, on the other hand, the mobility values in both molecular isomers are similar and are on the order of 10$^{-4}$ or smaller. 

Note that when QD size fluctuation (usually about 5$\%$) is introduced to the QD array, the difference in size between adjacent dots  changes the state energy difference $\epsilon_{a}-\epsilon_{b}$ of the QDs and hence the electron hopping rate $k_{ab}$.  In our systems, such a state energy fluctuation corresponds to about 50 meV\cite{doi:10.1021/jp206526s,PhysRevB.72.125325}. For each QD size, as shown in Figure \ref{rate_eab}, the electron hopping rates within this given energy range do not have any qualitative changes between the two isomer cases. Also, the carrier mobility with the QD size fluctuation should be of the same order of magnitude as the one computed before~\cite{doi:10.1021/jp206526s}, thus our conclusions drawn above should not change when considering a QD array with a distribution of dot sizes.

\section{Conclusions}
To summarize, we have studied electron hopping in a model system that contains two CdSe QDs linked by an AB derivative.  We have investigated the effects of molecular attachment, molecular isomer (\textit{trans} and \textit{cis}) and the QD size on the hopping rate. We have found that the molecular attachment has a strong impact on the system with both the isomers. In particular, the more energetically favorable attachment type II in the \textit{cis} isomer gives much higher QD-QD couplings than in the \textit{trans} isomer. This is due to the shorter QD-QD distance in the \textit{cis} isomer. Note, in isolation, the \textit{trans} isomer is more stable than the \textit{cis} isomer. We have also calculated the carrier mobility in the low carrier density regime in the corresponding QD supercrystal. When the more stable attachment type II is used, the calculated mobility values from the molecular \textit{trans} and \textit{cis} isomer differ by 500 times, which indicates that a molecular switch in such a QD supercrystal is feasible. 

Our results have demonstrated the first conductance switching device of its kind using QDs. Since the \textit{cis} configuration is more stable, one can also apply mechanical tension (in addition to light) to control the transition from the \textit{cis} configuration to the \textit{trans}. Results regarding the effect of attachment type on mobility are important to guide synthesis of such arrays to maximize differences between the $\textit{trans}$ and $\textit{cis}$ configurations. Our promising results suggest future experimental studies  to confirm the theoretical findings presented here.

\section{Acknowledgements}
This work is supported by the U.S. Department of Energy (DOE), Office of Basic Energy Sciences (BES) under Contract No. DE-FG02-02ER45995. L.W. Wang acknowledges the support by the Director, Office of Science (OS), BES/Materials Science and Engineering (MSED) of the U.S. DOE under the Contract No. DE-AC02-05CH11231. The calculations have been performed at NERSC and UF-HPC Center.


\begin{thebibliography}{44}
\expandafter\ifx\csname natexlab\endcsname\relax\def\natexlab#1{#1}\fi
\expandafter\ifx\csname bibnamefont\endcsname\relax
  \def\bibnamefont#1{#1}\fi
\expandafter\ifx\csname bibfnamefont\endcsname\relax
  \def\bibfnamefont#1{#1}\fi
\expandafter\ifx\csname citenamefont\endcsname\relax
  \def\citenamefont#1{#1}\fi
\expandafter\ifx\csname url\endcsname\relax
  \def\url#1{\texttt{#1}}\fi
\expandafter\ifx\csname urlprefix\endcsname\relax\def\urlprefix{URL }\fi
\providecommand{\bibinfo}[2]{#2}
\providecommand{\eprint}[2][]{\url{#2}}

\bibitem[{\citenamefont{Talapin et~al.}(2009)\citenamefont{Talapin, Lee,
  Kovalenko, and Shevchenko}}]{talapin2009prospects}
\bibinfo{author}{\bibfnamefont{D.~V.} \bibnamefont{Talapin}},
  \bibinfo{author}{\bibfnamefont{J.-S.} \bibnamefont{Lee}},
  \bibinfo{author}{\bibfnamefont{M.~V.} \bibnamefont{Kovalenko}},
  \bibnamefont{and} \bibinfo{author}{\bibfnamefont{E.~V.}
  \bibnamefont{Shevchenko}}, \bibinfo{journal}{Chem. Rev.}
  \textbf{\bibinfo{volume}{110}}, \bibinfo{pages}{389} (\bibinfo{year}{2009}).

\bibitem[{\citenamefont{Gur et~al.}(2005)\citenamefont{Gur, Fromer, Geier, and
  Alivisatos}}]{Gur21102005}
\bibinfo{author}{\bibfnamefont{I.}~\bibnamefont{Gur}},
  \bibinfo{author}{\bibfnamefont{N.~A.} \bibnamefont{Fromer}},
  \bibinfo{author}{\bibfnamefont{M.~L.} \bibnamefont{Geier}}, \bibnamefont{and}
  \bibinfo{author}{\bibfnamefont{A.~P.} \bibnamefont{Alivisatos}},
  \bibinfo{journal}{Science} \textbf{\bibinfo{volume}{310}},
  \bibinfo{pages}{462} (\bibinfo{year}{2005}).

\bibitem[{\citenamefont{Bendall et~al.}(2010)\citenamefont{Bendall, Paderi,
  Ghigliotti, Li~Pira, Lambertini, Lesnyak, Gaponik, Visimberga, Eychmüller,
  Torres et~al.}}]{bendall2010}
\bibinfo{author}{\bibfnamefont{J.~S.} \bibnamefont{Bendall}},
  \bibinfo{author}{\bibfnamefont{M.}~\bibnamefont{Paderi}},
  \bibinfo{author}{\bibfnamefont{F.}~\bibnamefont{Ghigliotti}},
  \bibinfo{author}{\bibfnamefont{N.}~\bibnamefont{Li~Pira}},
  \bibinfo{author}{\bibfnamefont{V.}~\bibnamefont{Lambertini}},
  \bibinfo{author}{\bibfnamefont{V.}~\bibnamefont{Lesnyak}},
  \bibinfo{author}{\bibfnamefont{N.}~\bibnamefont{Gaponik}},
  \bibinfo{author}{\bibfnamefont{G.}~\bibnamefont{Visimberga}},
  \bibinfo{author}{\bibfnamefont{A.}~\bibnamefont{Eychmüller}},
  \bibinfo{author}{\bibfnamefont{C.~M.~S.} \bibnamefont{Torres}},
  \bibnamefont{et~al.}, \bibinfo{journal}{Adv. Funct. Mater.}
  \textbf{\bibinfo{volume}{20}}, \bibinfo{pages}{3298} (\bibinfo{year}{2010}).

\bibitem[{\citenamefont{Wood and Bulovi{\'c}}(2010)}]{wood2010colloidal}
\bibinfo{author}{\bibfnamefont{V.}~\bibnamefont{Wood}} \bibnamefont{and}
  \bibinfo{author}{\bibfnamefont{V.}~\bibnamefont{Bulovi{\'c}}},
  \bibinfo{journal}{Nano Rev.} \textbf{\bibinfo{volume}{1}}
  (\bibinfo{year}{2010}).

\bibitem[{\citenamefont{Wang et~al.}(2008)\citenamefont{Wang, Feser, Lee,
  Talapin, Segalman, and Majumdar}}]{wang_thermopower}
\bibinfo{author}{\bibfnamefont{R.~Y.} \bibnamefont{Wang}},
  \bibinfo{author}{\bibfnamefont{J.~P.} \bibnamefont{Feser}},
  \bibinfo{author}{\bibfnamefont{J.-S.} \bibnamefont{Lee}},
  \bibinfo{author}{\bibfnamefont{D.~V.} \bibnamefont{Talapin}},
  \bibinfo{author}{\bibfnamefont{R.}~\bibnamefont{Segalman}}, \bibnamefont{and}
  \bibinfo{author}{\bibfnamefont{A.}~\bibnamefont{Majumdar}},
  \bibinfo{journal}{Nano Lett.} \textbf{\bibinfo{volume}{8}},
  \bibinfo{pages}{2283} (\bibinfo{year}{2008}).

\bibitem[{\citenamefont{Konstantatos et~al.}(2006)\citenamefont{Konstantatos,
  Howard, Fischer, Hoogland, Clifford, Klem, Levina, and
  Sargent}}]{konstantatos2006ultrasensitive}
\bibinfo{author}{\bibfnamefont{G.}~\bibnamefont{Konstantatos}},
  \bibinfo{author}{\bibfnamefont{I.}~\bibnamefont{Howard}},
  \bibinfo{author}{\bibfnamefont{A.}~\bibnamefont{Fischer}},
  \bibinfo{author}{\bibfnamefont{S.}~\bibnamefont{Hoogland}},
  \bibinfo{author}{\bibfnamefont{J.}~\bibnamefont{Clifford}},
  \bibinfo{author}{\bibfnamefont{E.}~\bibnamefont{Klem}},
  \bibinfo{author}{\bibfnamefont{L.}~\bibnamefont{Levina}}, \bibnamefont{and}
  \bibinfo{author}{\bibfnamefont{E.~H.} \bibnamefont{Sargent}},
  \bibinfo{journal}{Nature} \textbf{\bibinfo{volume}{442}},
  \bibinfo{pages}{180} (\bibinfo{year}{2006}).

\bibitem[{\citenamefont{Yu et~al.}(2003)\citenamefont{Yu, Wang, and
  Guyot-Sionnest}}]{Yu23052003}
\bibinfo{author}{\bibfnamefont{D.}~\bibnamefont{Yu}},
  \bibinfo{author}{\bibfnamefont{C.}~\bibnamefont{Wang}}, \bibnamefont{and}
  \bibinfo{author}{\bibfnamefont{P.}~\bibnamefont{Guyot-Sionnest}},
  \bibinfo{journal}{Science} \textbf{\bibinfo{volume}{300}},
  \bibinfo{pages}{1277} (\bibinfo{year}{2003}).

\bibitem[{\citenamefont{Alivisatos}(1996)}]{Alivisatos16021996}
\bibinfo{author}{\bibfnamefont{A.~P.} \bibnamefont{Alivisatos}},
  \bibinfo{journal}{Science} \textbf{\bibinfo{volume}{271}},
  \bibinfo{pages}{933} (\bibinfo{year}{1996}).

\bibitem[{\citenamefont{Vanmaekelbergh and
  Liljeroth}(2005)}]{vanmaekelbergh2005electron}
\bibinfo{author}{\bibfnamefont{D.}~\bibnamefont{Vanmaekelbergh}}
  \bibnamefont{and}
  \bibinfo{author}{\bibfnamefont{P.}~\bibnamefont{Liljeroth}},
  \bibinfo{journal}{Chem. Soc. Rev.} \textbf{\bibinfo{volume}{34}},
  \bibinfo{pages}{299} (\bibinfo{year}{2005}).

\bibitem[{\citenamefont{Morgan et~al.}(2002)\citenamefont{Morgan, Leatherdale,
  Drndi\ifmmode~\acute{c}\else \'{c}\fi{}, Jarosz, Kastner, and
  Bawendi}}]{morganPhysRevB.66.075339}
\bibinfo{author}{\bibfnamefont{N.~Y.} \bibnamefont{Morgan}},
  \bibinfo{author}{\bibfnamefont{C.~A.} \bibnamefont{Leatherdale}},
  \bibinfo{author}{\bibfnamefont{M.}~\bibnamefont{Drndi\ifmmode~\acute{c}\else
  \'{c}\fi{}}}, \bibinfo{author}{\bibfnamefont{M.~V.} \bibnamefont{Jarosz}},
  \bibinfo{author}{\bibfnamefont{M.~A.} \bibnamefont{Kastner}},
  \bibnamefont{and} \bibinfo{author}{\bibfnamefont{M.}~\bibnamefont{Bawendi}},
  \bibinfo{journal}{Phys. Rev. B} \textbf{\bibinfo{volume}{66}},
  \bibinfo{pages}{075339} (\bibinfo{year}{2002}).

\bibitem[{\citenamefont{Lee et~al.}(2011)\citenamefont{Lee, Kovalenko, Huang,
  Chung, and Talapin}}]{lee2011band}
\bibinfo{author}{\bibfnamefont{J.-S.} \bibnamefont{Lee}},
  \bibinfo{author}{\bibfnamefont{M.~V.} \bibnamefont{Kovalenko}},
  \bibinfo{author}{\bibfnamefont{J.}~\bibnamefont{Huang}},
  \bibinfo{author}{\bibfnamefont{D.~S.} \bibnamefont{Chung}}, \bibnamefont{and}
  \bibinfo{author}{\bibfnamefont{D.~V.} \bibnamefont{Talapin}},
  \bibinfo{journal}{Nature Nanotechnol.} \textbf{\bibinfo{volume}{6}},
  \bibinfo{pages}{348} (\bibinfo{year}{2011}).

\bibitem[{\citenamefont{Hugel et~al.}(2002)\citenamefont{Hugel, Holland,
  Cattani, Moroder, Seitz, and Gaub}}]{Hugel10052002}
\bibinfo{author}{\bibfnamefont{T.}~\bibnamefont{Hugel}},
  \bibinfo{author}{\bibfnamefont{N.~B.} \bibnamefont{Holland}},
  \bibinfo{author}{\bibfnamefont{A.}~\bibnamefont{Cattani}},
  \bibinfo{author}{\bibfnamefont{L.}~\bibnamefont{Moroder}},
  \bibinfo{author}{\bibfnamefont{M.}~\bibnamefont{Seitz}}, \bibnamefont{and}
  \bibinfo{author}{\bibfnamefont{H.~E.} \bibnamefont{Gaub}},
  \bibinfo{journal}{Science} \textbf{\bibinfo{volume}{296}},
  \bibinfo{pages}{1103} (\bibinfo{year}{2002}).

\bibitem[{\citenamefont{Zhang et~al.}(2004)\citenamefont{Zhang, Du, Cheng,
  Zhang, Roitberg, and Krause}}]{PhysRevLett.92.158301}
\bibinfo{author}{\bibfnamefont{C.}~\bibnamefont{Zhang}},
  \bibinfo{author}{\bibfnamefont{M.-H.} \bibnamefont{Du}},
  \bibinfo{author}{\bibfnamefont{H.-P.} \bibnamefont{Cheng}},
  \bibinfo{author}{\bibfnamefont{X.-G.} \bibnamefont{Zhang}},
  \bibinfo{author}{\bibfnamefont{A.~E.} \bibnamefont{Roitberg}},
  \bibnamefont{and} \bibinfo{author}{\bibfnamefont{J.~L.}
  \bibnamefont{Krause}}, \bibinfo{journal}{Phys. Rev. Lett.}
  \textbf{\bibinfo{volume}{92}}, \bibinfo{pages}{158301}
  (\bibinfo{year}{2004}).

\bibitem[{\citenamefont{Del~Valle et~al.}(2007)\citenamefont{Del~Valle,
  Gutierrez, Tejedor, and Cuniberti}}]{del2007tuning}
\bibinfo{author}{\bibfnamefont{M.}~\bibnamefont{Del~Valle}},
  \bibinfo{author}{\bibfnamefont{R.}~\bibnamefont{Gutierrez}},
  \bibinfo{author}{\bibfnamefont{C.}~\bibnamefont{Tejedor}}, \bibnamefont{and}
  \bibinfo{author}{\bibfnamefont{G.}~\bibnamefont{Cuniberti}},
  \bibinfo{journal}{Nature Nanotechnol.} \textbf{\bibinfo{volume}{2}},
  \bibinfo{pages}{176} (\bibinfo{year}{2007}).

\bibitem[{\citenamefont{Wang and Cheng}(2012)}]{yw-azob}
\bibinfo{author}{\bibfnamefont{Y.}~\bibnamefont{Wang}} \bibnamefont{and}
  \bibinfo{author}{\bibfnamefont{H.-P.} \bibnamefont{Cheng}},
  \bibinfo{journal}{Phys. Rev. B} \textbf{\bibinfo{volume}{86}},
  \bibinfo{pages}{035444} (\bibinfo{year}{2012}).

\bibitem[{\citenamefont{Hagen and Bieringer}(2001)}]{hagen2001photoaddressable}
\bibinfo{author}{\bibfnamefont{R.}~\bibnamefont{Hagen}} \bibnamefont{and}
  \bibinfo{author}{\bibfnamefont{T.}~\bibnamefont{Bieringer}},
  \bibinfo{journal}{Adv. Mater.} \textbf{\bibinfo{volume}{13}},
  \bibinfo{pages}{1805} (\bibinfo{year}{2001}).

\bibitem[{\citenamefont{Mativetsky et~al.}(2008)\citenamefont{Mativetsky, Pace,
  Elbing, Rampi, Mayor, and Samorì}}]{mativetsky_azo}
\bibinfo{author}{\bibfnamefont{J.~M.} \bibnamefont{Mativetsky}},
  \bibinfo{author}{\bibfnamefont{G.}~\bibnamefont{Pace}},
  \bibinfo{author}{\bibfnamefont{M.}~\bibnamefont{Elbing}},
  \bibinfo{author}{\bibfnamefont{M.~A.} \bibnamefont{Rampi}},
  \bibinfo{author}{\bibfnamefont{M.}~\bibnamefont{Mayor}}, \bibnamefont{and}
  \bibinfo{author}{\bibfnamefont{P.}~\bibnamefont{Samorì}},
  \bibinfo{journal}{J. Am. Chem. Soc.} \textbf{\bibinfo{volume}{130}},
  \bibinfo{pages}{9192} (\bibinfo{year}{2008}).

\bibitem[{\citenamefont{Fang et~al.}(2010)\citenamefont{Fang, Chen, Liu, Zhang,
  and Ding}}]{fang2010temperature}
\bibinfo{author}{\bibfnamefont{D.}~\bibnamefont{Fang}},
  \bibinfo{author}{\bibfnamefont{R.}~\bibnamefont{Chen}},
  \bibinfo{author}{\bibfnamefont{H.}~\bibnamefont{Liu}},
  \bibinfo{author}{\bibfnamefont{Z.}~\bibnamefont{Zhang}}, \bibnamefont{and}
  \bibinfo{author}{\bibfnamefont{Z.}~\bibnamefont{Ding}}, \bibinfo{journal}{J.
  Nanosci. Nanotechnol.} \textbf{\bibinfo{volume}{10}}, \bibinfo{pages}{7600}
  (\bibinfo{year}{2010}).

\bibitem[{\citenamefont{Klajn et~al.}(2007)\citenamefont{Klajn, Bishop, and
  Grzybowski}}]{Klajn19062007}
\bibinfo{author}{\bibfnamefont{R.}~\bibnamefont{Klajn}},
  \bibinfo{author}{\bibfnamefont{K.~J.~M.} \bibnamefont{Bishop}},
  \bibnamefont{and} \bibinfo{author}{\bibfnamefont{B.~A.}
  \bibnamefont{Grzybowski}}, \bibinfo{journal}{Proc. Natl. Acad. Sci.}
  \textbf{\bibinfo{volume}{104}}, \bibinfo{pages}{10305}
  (\bibinfo{year}{2007}).

\bibitem[{\citenamefont{Chu et~al.}(2011)\citenamefont{Chu, Radulaski,
  Vukmirovic, Cheng, and Wang}}]{doi:10.1021/jp206526s}
\bibinfo{author}{\bibfnamefont{I.-H.} \bibnamefont{Chu}},
  \bibinfo{author}{\bibfnamefont{M.}~\bibnamefont{Radulaski}},
  \bibinfo{author}{\bibfnamefont{N.}~\bibnamefont{Vukmirovic}},
  \bibinfo{author}{\bibfnamefont{H.-P.} \bibnamefont{Cheng}}, \bibnamefont{and}
  \bibinfo{author}{\bibfnamefont{L.-W.} \bibnamefont{Wang}},
  \bibinfo{journal}{The Journal of Physical Chemistry C}
  \textbf{\bibinfo{volume}{115}}, \bibinfo{pages}{21409}
  (\bibinfo{year}{2011}).

\bibitem[{\citenamefont{Marcus}(1993)}]{marcus}
\bibinfo{author}{\bibfnamefont{R.~A.} \bibnamefont{Marcus}},
  \bibinfo{journal}{Rev. Mod. Phys.} \textbf{\bibinfo{volume}{65}},
  \bibinfo{pages}{599} (\bibinfo{year}{1993}).

\bibitem[{\citenamefont{Kilina et~al.}(2011)\citenamefont{Kilina, Kilin,
  Prezhdo, and Prezhdo}}]{relax1}
\bibinfo{author}{\bibfnamefont{S.~V.} \bibnamefont{Kilina}},
  \bibinfo{author}{\bibfnamefont{D.~S.} \bibnamefont{Kilin}},
  \bibinfo{author}{\bibfnamefont{V.~V.} \bibnamefont{Prezhdo}},
  \bibnamefont{and} \bibinfo{author}{\bibfnamefont{O.~V.}
  \bibnamefont{Prezhdo}}, \bibinfo{journal}{J. Phys. Chem. C}
  \textbf{\bibinfo{volume}{115}}, \bibinfo{pages}{21641}
  (\bibinfo{year}{2011}).

\bibitem[{\citenamefont{Guyot-Sionnest
  et~al.}(1999)\citenamefont{Guyot-Sionnest, Shim, Matranga, and
  Hines}}]{relax2}
\bibinfo{author}{\bibfnamefont{P.}~\bibnamefont{Guyot-Sionnest}},
  \bibinfo{author}{\bibfnamefont{M.}~\bibnamefont{Shim}},
  \bibinfo{author}{\bibfnamefont{C.}~\bibnamefont{Matranga}}, \bibnamefont{and}
  \bibinfo{author}{\bibfnamefont{M.}~\bibnamefont{Hines}},
  \bibinfo{journal}{Phys. Rev. B} \textbf{\bibinfo{volume}{60}},
  \bibinfo{pages}{R2181} (\bibinfo{year}{1999}).

\bibitem[{\citenamefont{Kovalenko et~al.}(2009)\citenamefont{Kovalenko,
  Scheele, and Talapin}}]{Kovalenko12062009}
\bibinfo{author}{\bibfnamefont{M.~V.} \bibnamefont{Kovalenko}},
  \bibinfo{author}{\bibfnamefont{M.}~\bibnamefont{Scheele}}, \bibnamefont{and}
  \bibinfo{author}{\bibfnamefont{D.~V.} \bibnamefont{Talapin}},
  \bibinfo{journal}{Science} \textbf{\bibinfo{volume}{324}},
  \bibinfo{pages}{1417} (\bibinfo{year}{2009}).

\bibitem[{xcr()}]{xcrysden}
\bibinfo{howpublished}{\url{http://www.xcrysden.org}}.

\bibitem[{\citenamefont{Manna et~al.}(2005)\citenamefont{Manna, Wang,
  Cingolani, and Alivisatos}}]{doi:10.1021/jp0445573}
\bibinfo{author}{\bibfnamefont{L.}~\bibnamefont{Manna}},
  \bibinfo{author}{\bibnamefont{Wang}},
  \bibinfo{author}{\bibfnamefont{R.}~\bibnamefont{Cingolani}},
  \bibnamefont{and} \bibinfo{author}{\bibfnamefont{A.~P.}
  \bibnamefont{Alivisatos}}, \bibinfo{journal}{J. Phys. Chem. B}
  \textbf{\bibinfo{volume}{109}}, \bibinfo{pages}{6183} (\bibinfo{year}{2005}).

\bibitem[{\citenamefont{Kresse and Hafner}(1993)}]{PhysRevB.47.558}
\bibinfo{author}{\bibfnamefont{G.}~\bibnamefont{Kresse}} \bibnamefont{and}
  \bibinfo{author}{\bibfnamefont{J.}~\bibnamefont{Hafner}},
  \bibinfo{journal}{Phys. Rev. B} \textbf{\bibinfo{volume}{47}},
  \bibinfo{pages}{558} (\bibinfo{year}{1993}).

\bibitem[{\citenamefont{Kresse and Furthm\"uller}(1996)}]{PhysRevB.54.11169}
\bibinfo{author}{\bibfnamefont{G.}~\bibnamefont{Kresse}} \bibnamefont{and}
  \bibinfo{author}{\bibfnamefont{J.}~\bibnamefont{Furthm\"uller}},
  \bibinfo{journal}{Phys. Rev. B} \textbf{\bibinfo{volume}{54}},
  \bibinfo{pages}{11169} (\bibinfo{year}{1996}).

\bibitem[{\citenamefont{Ceperley and Alder}(1980)}]{PhysRevLett.45.566}
\bibinfo{author}{\bibfnamefont{D.~M.} \bibnamefont{Ceperley}} \bibnamefont{and}
  \bibinfo{author}{\bibfnamefont{B.~J.} \bibnamefont{Alder}},
  \bibinfo{journal}{Phys. Rev. Lett.} \textbf{\bibinfo{volume}{45}},
  \bibinfo{pages}{566} (\bibinfo{year}{1980}).

\bibitem[{\citenamefont{Bl\"ochl}(1994)}]{PhysRevB.50.17953}
\bibinfo{author}{\bibfnamefont{P.~E.} \bibnamefont{Bl\"ochl}},
  \bibinfo{journal}{Phys. Rev. B} \textbf{\bibinfo{volume}{50}},
  \bibinfo{pages}{17953} (\bibinfo{year}{1994}).

\bibitem[{pet()}]{petot}
\bibinfo{howpublished}{\url{http://cmsn.lbl.gov/html/PEtot/PEtot.html}}.

\bibitem[{\citenamefont{Troullier and Martins}(1991)}]{NCPP}
\bibinfo{author}{\bibfnamefont{N.}~\bibnamefont{Troullier}} \bibnamefont{and}
  \bibinfo{author}{\bibfnamefont{J.~L.} \bibnamefont{Martins}},
  \bibinfo{journal}{Phys. Rev. B} \textbf{\bibinfo{volume}{43}},
  \bibinfo{pages}{1993} (\bibinfo{year}{1991}).

\bibitem[{\citenamefont{Wang}(2002)}]{PhysRevLett.88.256402}
\bibinfo{author}{\bibfnamefont{L.-W.} \bibnamefont{Wang}},
  \bibinfo{journal}{Phys. Rev. Lett.} \textbf{\bibinfo{volume}{88}},
  \bibinfo{pages}{256402} (\bibinfo{year}{2002}).

\bibitem[{\citenamefont{Kohn and Sham}(1965)}]{KS}
\bibinfo{author}{\bibfnamefont{W.}~\bibnamefont{Kohn}} \bibnamefont{and}
  \bibinfo{author}{\bibfnamefont{L.~J.} \bibnamefont{Sham}},
  \bibinfo{journal}{Phys. Rev.} \textbf{\bibinfo{volume}{140}},
  \bibinfo{pages}{A1133} (\bibinfo{year}{1965}).

\bibitem[{\citenamefont{Wang and Zunger}(1994)}]{FSM}
\bibinfo{author}{\bibfnamefont{L.}~\bibnamefont{Wang}} \bibnamefont{and}
  \bibinfo{author}{\bibfnamefont{A.}~\bibnamefont{Zunger}},
  \bibinfo{journal}{J. Chem. Phys.} \textbf{\bibinfo{volume}{100}},
  \bibinfo{pages}{2394} (\bibinfo{year}{1994}).

\bibitem[{\citenamefont{Wang}(1994)}]{GMM}
\bibinfo{author}{\bibfnamefont{L.-W.} \bibnamefont{Wang}},
  \bibinfo{journal}{Phys. Rev. B} \textbf{\bibinfo{volume}{49}},
  \bibinfo{pages}{10154} (\bibinfo{year}{1994}).

\bibitem[{\citenamefont{Shan et~al.}(1999)\citenamefont{Shan, Walukiewicz,
  Ager, Haller, Geisz, Friedman, Olson, and Kurtz}}]{AC}
\bibinfo{author}{\bibfnamefont{W.}~\bibnamefont{Shan}},
  \bibinfo{author}{\bibfnamefont{W.}~\bibnamefont{Walukiewicz}},
  \bibinfo{author}{\bibfnamefont{J.~W.} \bibnamefont{Ager}},
  \bibinfo{author}{\bibfnamefont{E.~E.} \bibnamefont{Haller}},
  \bibinfo{author}{\bibfnamefont{J.~F.} \bibnamefont{Geisz}},
  \bibinfo{author}{\bibfnamefont{D.~J.} \bibnamefont{Friedman}},
  \bibinfo{author}{\bibfnamefont{J.~M.} \bibnamefont{Olson}}, \bibnamefont{and}
  \bibinfo{author}{\bibfnamefont{S.~R.} \bibnamefont{Kurtz}},
  \bibinfo{journal}{Phys. Rev. Lett.} \textbf{\bibinfo{volume}{82}},
  \bibinfo{pages}{1221} (\bibinfo{year}{1999}).

\bibitem[{\citenamefont{Huang and Rhys}(1950)}]{Huang22121950}
\bibinfo{author}{\bibfnamefont{K.}~\bibnamefont{Huang}} \bibnamefont{and}
  \bibinfo{author}{\bibfnamefont{A.}~\bibnamefont{Rhys}},
  \bibinfo{journal}{Proc. R. Soc. A} \textbf{\bibinfo{volume}{204}},
  \bibinfo{pages}{406} (\bibinfo{year}{1950}).

\bibitem[{\citenamefont{Nan et~al.}(2009)\citenamefont{Nan, Yang, Wang, Shuai,
  and Zhao}}]{PhysRevB.79.115203}
\bibinfo{author}{\bibfnamefont{G.}~\bibnamefont{Nan}},
  \bibinfo{author}{\bibfnamefont{X.}~\bibnamefont{Yang}},
  \bibinfo{author}{\bibfnamefont{L.}~\bibnamefont{Wang}},
  \bibinfo{author}{\bibfnamefont{Z.}~\bibnamefont{Shuai}}, \bibnamefont{and}
  \bibinfo{author}{\bibfnamefont{Y.}~\bibnamefont{Zhao}},
  \bibinfo{journal}{Phys. Rev. B} \textbf{\bibinfo{volume}{79}},
  \bibinfo{pages}{115203} (\bibinfo{year}{2009}).

\bibitem[{\citenamefont{Keating}(1966)}]{VFF}
\bibinfo{author}{\bibfnamefont{P.~N.} \bibnamefont{Keating}},
  \bibinfo{journal}{Phys. Rev.} \textbf{\bibinfo{volume}{145}},
  \bibinfo{pages}{637} (\bibinfo{year}{1966}).

\bibitem[{\citenamefont{Ambegaokar et~al.}(1971)\citenamefont{Ambegaokar,
  Halperin, and Langer}}]{PhysRevB.4.2612}
\bibinfo{author}{\bibfnamefont{V.}~\bibnamefont{Ambegaokar}},
  \bibinfo{author}{\bibfnamefont{B.~I.} \bibnamefont{Halperin}},
  \bibnamefont{and} \bibinfo{author}{\bibfnamefont{J.~S.}
  \bibnamefont{Langer}}, \bibinfo{journal}{Phys. Rev. B}
  \textbf{\bibinfo{volume}{4}}, \bibinfo{pages}{2612} (\bibinfo{year}{1971}).

\bibitem[{\citenamefont{Vukmirović and Wang}(2009)}]{doi:10.1021/nl9021539}
\bibinfo{author}{\bibfnamefont{N.}~\bibnamefont{Vukmirović}}
  \bibnamefont{and} \bibinfo{author}{\bibfnamefont{L.-W.} \bibnamefont{Wang}},
  \bibinfo{journal}{Nano Lett.} \textbf{\bibinfo{volume}{9}},
  \bibinfo{pages}{3996} (\bibinfo{year}{2009}).

\bibitem[{\citenamefont{Coropceanu et~al.}(2007)\citenamefont{Coropceanu,
  Cornil, da~Silva~Filho, Olivier, Silbey, and Brédas}}]{transport_review}
\bibinfo{author}{\bibfnamefont{V.}~\bibnamefont{Coropceanu}},
  \bibinfo{author}{\bibfnamefont{J.}~\bibnamefont{Cornil}},
  \bibinfo{author}{\bibfnamefont{D.~A.} \bibnamefont{da~Silva~Filho}},
  \bibinfo{author}{\bibfnamefont{Y.}~\bibnamefont{Olivier}},
  \bibinfo{author}{\bibfnamefont{R.}~\bibnamefont{Silbey}}, \bibnamefont{and}
  \bibinfo{author}{\bibfnamefont{J.-L.} \bibnamefont{Brédas}},
  \bibinfo{journal}{Chem. Rev.} \textbf{\bibinfo{volume}{107}},
  \bibinfo{pages}{926} (\bibinfo{year}{2007}).

\bibitem[{\citenamefont{Li and Wang}(2005)}]{PhysRevB.72.125325}
\bibinfo{author}{\bibfnamefont{J.}~\bibnamefont{Li}} \bibnamefont{and}
  \bibinfo{author}{\bibfnamefont{L.-W.} \bibnamefont{Wang}},
  \bibinfo{journal}{Phys. Rev. B} \textbf{\bibinfo{volume}{72}},
  \bibinfo{pages}{125325} (\bibinfo{year}{2005}).

\end{thebibliography}
\end{document}